\documentclass[amssymb,aps,twocolumn,prb]{revtex4-1}
\usepackage{epsfig,subfigure,amsmath,float,xcolor,ulem}
\usepackage[colorlinks]{hyperref}
\hypersetup{colorlinks=true,linkcolor=blue,filecolor=blue,urlcolor=blue,citecolor=blue}
\usepackage[utf8]{inputenc}
\usepackage{soul}

\newcommand\beq{\begin{equation}}
\newcommand\eeq{\end{equation}}
\newcommand\bes{\begin{subequations}}
\newcommand\ees{\end{subequations}}
\newcommand\bea{\begin{eqnarray}}
\newcommand\eea{\end{eqnarray}}
\newcommand\non{\nonumber}
\newcommand\eps{\epsilon}

\newcommand\ig{\includegraphics}

\newcommand\al{\alpha}
\usepackage{yfonts}
\newcommand\be{\beta}

\newcommand\la{\langle}
\newcommand\ra{\rangle}
\newcommand\om{\omega}

\newcommand\Lam{\Lambda}

\definecolor{lilac}{rgb}{0.4, 0.5 ,1}
\begin{document}
\title{Response of the Quantum Ground State to a Parametric Drive}
\author{Ranjani Seshadri}
\email{seshad@bc.edu}
\affiliation{Department of Physics, Boston College, 140 Commonwealth Avenue Chestnut Hill,
Massachusetts 02467, USA}
\date{\today}
\begin{abstract}
The phenomenon of Parametric Resonance (PR) is very well studied in classical systems with
one of the textbook examples being the stabilization of a Kapitza's pendulum in the inverted
configuration when the suspension point is oscillated vertically.  One important aspect that
distinguishes between classical PR and ordinary resonance is that in the former, if the initial
energy of the system is at its minimum (${\dot x}={x}=0$), the system does not evolve. In a
quantum system, however, even when the system is in the minimum energy (ground) state, the
system has non-trivial evolution under PR due to the delocalized nature of the ground state
wavefunction. Here we study the evolution of such a system which exhibits a purely quantum effect
with no classical analog. In particular, we focus on the quantum mechanical analog of PR by
varying with time the parabolic potential i.e. the frequency of the quantum harmonic oscillator. 
\end{abstract}

\maketitle
\section{Introduction} 


The study of driven quantum systems has been a very active area of research. 
One way to realize such a time-dependent system is by choosing a parameter
in the equilibrium Hamiltonian and varying it with time. In this context we study the
simple yet ubiquitous quantum harmonic oscillator  - one of the cornerstones of quantum
mechanics - with a time dependence in the "frequency". In classical systems,
under special circumstances, this can exhibit the phenomenon of parametric resonance
(PR) which is very well understood.
The most well known example is Kapitza's pendulum where the pivot point of the pendulum
is oscillated vertically. {{The unique feature of this pendulum is the stabilization
in the inverted configuration which is explained by PR. This is a point of unstable equilibrium,
where the bob of the pendulum is above the pivot point, and is not found in the usual pendulum
with a fixed suspension.}}

The classical equation of motion governing the dynamics of a parametrically driven oscillator
is given by

\begin{equation}
{\ddot x}+\omega_0^2f(t)x=0 \label{eq:HO}
\end{equation}
where, $w_0$ is the natural frequency of the pendulum. We choose the specific form of the
perturbation $f(t) = (1+h \sin(({2 \om_0+\eps})t)$ where the amplitude $h$ and $\eps$ are the
drive parameters.


It is a well-known result \cite{landau} that classical PR occurs for the case when 
\beq
|\epsilon| < \frac{1}{2}h \omega_0 \label{eq:eh}
\eeq
In this classical case, if the initial condition of the system is assumed to be the minimum energy
state, i.e.  $x(t_{i})={\dot x}(t_{i})=0$, then the system never evolves and continues to remain
in the initial state. This is the most important aspect that distinguishes ordinary resonance from
PR. In terms of the Kapitza pendulum, this can be understood as follows. When the
suspension point of a simple pendulum is periodically oscillated vertically, the phenomenon of
PR is observed only if the bob of the pendulum has a non-zero displacement
or a non-zero velocity to begin with. If the bob of the pendulum is at the energy  minimum,
it remains undisturbed by this perturbation.
\cite{akridge95, weig2002}

The quantum mechanical counterpart, however, exhibits a markedly different behaviour.
In the minimum energy (i.e. ground state) configuration the wave function, although peaked
at the potential minimum (assumed to be at $x=0$), is not localized at $x=0$. Therefore, the
dynamics, even in the ground state, is affected by the time-dependent perturbation $f(t)$ in
Eq. \eqref{eq:HO}. In particular,
the conditions of PR in the quantum treatment of the particle in the ground
state results in effects which are purely quantum in nature with no classical analogy. 
We note that the phenomenon of parametric resonance has been investigated in other quantum
mechanical systems, for example in phonon in an irradiated quantum well \cite{phonons},
quantum electrodynamics \cite{qprfield} and in quantum zeno effect \cite{zeno}.
The purpose of this paper is to investigate these exclusive quantum effects on the ground state
of a harmonic oscillator
under the conditions that correspond to that of PR at the classical level.

\section{Parametrically Driven Quantum Harmonic Oscillator}
The evolution of the wavefunction, $\phi \equiv \phi(x)$ of the  quantum harmonic
oscillator with mass $m$ and {{constant}} frequency $\om_0$ is governed by the 
time-independent Schro\"dinger Equation,
\beq
-\frac{\hbar^2}{2m} \frac{d^2 \phi}{d x^2} + \frac{1}{2}m\om_0^2x^2 \phi = E \phi. \non
\eeq
{In terms of the dimensionless position, $\xi =\sqrt{{m \om_0}/{\hbar}}~x$ and
the dimensionless energy, $\Lam = {2E}/{\hbar \om_0}$, this equation} reduces to the form
\beq
\frac{d^2 \psi}{d \xi^2}+ (\Lam - \xi^2)\psi = 0 \label{eq:SHO_tinda}.
\eeq
The $n$th eigenstate of the system {described} by Eq. \eqref{eq:SHO_tinda} is
\beq
\psi_n(\xi) = N_n H_n(\xi)e^{-{\xi^2}/{2}},\label{eq:psin0}
\eeq
with energy $E_n = (n+1/2)\hbar\omega_0$, $H_n(\xi)$ is the $n$th order Hermite
polynomial and $N_n$ (chosen to be real) is the normalization constant set
by the condition
\beq
\int_{-\infty}^{+\infty} \frac{d\xi}{\sqrt{{m \om_0}/{\hbar}}} ~~|\psi_n(\xi)|^2 = 1.
\eeq
{As in the classical case of Eq. \eqref{eq:HO} we now vary the frequency,
$\om(t)^2 = \om_0^2 f(t)$ where,
\bea
f(t)=\begin{cases}1+h\sin((2\om_0+\eps)t),~~~\text{if}~~~ 0<t<\frac{\nu\pi}{2\om_0+\eps},\\
1, ~~~~~~~~~~~~~~~~~~~~~~~~~~~~~~~~~~~~~~~ \text{otherwise}.
\end{cases}\label{eq:ft}
\eea
The form of $f(t)$ is chosen such that even though there is a sharp switch-on (at $t=t_i=0$)
and switch-off time (at $t=t_f = \frac{\nu\pi}{2\om_0+\eps})$, the potential is always
a continuous function of time, $\nu \in \mathbb{Z}^+$ being the number of drive cycles. The corresponding
time-dependent Schr\"odinger equation is
\beq
i \hbar \frac{\partial \Phi(x,t)}{\partial t} = -\frac{\hbar^2}{2m}
\frac{\partial^2\Phi(x,t)}{\partial x^2}+\frac{1}{2}m\om^2(t)x^2\Phi(x,t). \label{eq:SHO_tdd}
\eeq
In terms of the dimensionless position $\xi =\sqrt{{m \om_0}/{\hbar}}~x$ and
time $\tau = \om_0 t$ it is expressed as
\beq
i\frac{\partial\Psi(\xi,\tau)}{\partial\tau}=-\frac{1}{2}\frac{\partial^2\Psi(\xi,\tau)}{\partial\xi^2} 
+\frac{1}{2} \xi^2 g_{h,\bar{\eps}}(\tau) \Psi(\xi,\tau). \label{eq:SHO_tdnd}
\eeq
Here $g_{h,\bar{\eps}}(\tau)$ is the dimensionless equivalent of $f(t)$
\bea
g_{h,\bar{\eps}}(\tau)=\begin{cases}1+h\sin((2+\bar\eps)\tau),~~~\text{if}~~~0<\tau<\frac{\nu\pi 
}{2+\bar\eps},\\1, ~~~~~~~~~~~~~~~~~~~~~~~~~~~~~~~ \text{otherwise}
\end{cases}\label{eq:gtau}
\eea
 and $\bar\eps = \eps/\om_0$. 
 
{{\section{Time Evolution of ground state}}}
We study the case when the system is in the ground state for $\tau<0$.
The ground state wave function of the unperturbed oscillator can be obtained from Eq. \eqref{eq:psin0}
by setting $n=0$ and is known to be a Gaussian. We use the fact that a Gaussian wavefunction in a
quadratic potential, even if it is time-dependent, evolves into a Gaussian of a different width.
Therefore, the time-evolution of the initial ground state is given by,

\beq
\Psi(\xi,\tau) = \begin{cases} N_0 e^{-{{\xi^2}/{2}}}~~~~~~~~\text{if}~~~\tau\leq0\\
A(\tau)e^{-B(\tau)\xi^2} ~~~\text{if}~~~\tau>0,\end{cases} \label{eq:genGauss}
\eeq
where $A(\tau)$ and $B(\tau)$ are, in general, complex functions.

For the above wavefunction to be an acceptable solution, time-evolution has to be unitary
and hence the normalization condition $\int_{-\infty}^{+\infty}\frac{d\xi}{\sqrt{{m
\om_0}/{\hbar}}} |\Psi(\xi,\tau)|^2 = 1$ should be satisfied.

Using the Eq. \eqref{eq:genGauss} in Eq. \eqref{eq:SHO_tdnd}, and noting that it holds for
all $\xi$ we get
\bea
B-i\frac{\dot{A}}{A} &=& 0,~~~\text{and}  \non \\
\frac{1}{2}g_{h,\bar{\eps}}+i\dot{B}-2B^2&=&0.
\eea
Therefore, the real and imaginary parts of $A = A_R+iA_I$ and $B =B_R+iB_I$ satisfy the following 
set of coupled differential equations,

\bea
A_R B_R - A_I B_I + \dot{A}_I &=& 0 \non \\  
A_R B_I + A_I B_R - \dot{A}_R &=& 0 \non \\ 
g-2\dot{B}_I-4(B_R^2-B_I^2) &=& 0 \non \\
\dot{B}_R - 4 B_R B_I &=& 0
\eea

We numerically solve the above set of differential equations with the initial conditions
$A(0) = N_0$ and $B(0) = 1/2$, which corresponds to the ground state (at $\tau\le 0$) to obtain
$A(\tau_f)$ and $B(\tau_f)$ at the final time. We can then reconstruct the wave function
$\Psi(\xi,\tau)$ of the evolved state at $\tau_f$ using Eq. \eqref{eq:genGauss}. The results
are analysed in terms of $p_n$ which is the probability of the evolved state to be in the $n$th
eigenstate of the unperturbed oscillator i.e.,
\beq
p_n = |\la \psi_n(\xi)| \Psi (\xi,\tau_f) \ra|^2. \label{eq:pndef}
\eeq
This can, in general depend on all the parameters of the drive, namely  the amplitude $h$, the
number of drive cycles $\nu$, and $\bar{\eps}$. Of particular interest will be the case when the
time dependence satisfies the PR condition of Eq. \eqref{eq:eh}.

Before we go on to study the numerical results and the dependence of $p_n$ on the drive parameters,
we first make some preliminary observations. Since the time-evolved Gaussian $ \Psi (\xi,\tau_f) $
is {\it different} from the ground-state Gaussian $\psi_0(\xi)$, we can immediately infer that there
is, {\it in general}, a non-zero projection $\langle\psi_n(\xi)|\Psi (\xi,\tau_f)\rangle$ of the
evolved state on to the excited states (i.e. $n>0$). However, since the evolved state is still an
even function of position, the projection on to the odd $n$ excited states is identically zero i.e.,
\beq
p_n = 0,~~~\text{for odd n}.
\eeq

For convenience, we define a dimensionless parameter $r=\bar\eps/h$.
The classical PR condition given in Eq. \eqref{eq:eh} can be rewritten in terms
of this dimensionless parameter as
\beq
|r|<0.5.
\eeq
The behaviour of the probability $p_n$ can be analyzed as a function of $r$. Had we considered
states with non-zero mean, then the effect of classical PR will be expected to
affect the mean value according to Ehrenfest's theorem. In our case, of course, the mean values
continue to be zero and any effect we are studying due to time dependence of parameters, in
particular PR, is a pure quantum effect. 

The plot of $p_n$ as a function of $r$ for even values of $n$ $(= 0,2,4,6)$ is shown in
Fig. \ref{fig:Fig2} where each panel corresponds to a different pair of $(h,\nu)$. Before
quantifying the behaviour of $p_n$ vs $r$ for different values of $n$, we observe that
there is a clear transition close to $|r|=0.5$. This transition becomes sharper with
increasing number of oscillations $\nu$ and/or the amplitude $h$. When $|r|>0.5$, the
value of $p_n$ for $n=0$ seems to dominate i.e. $p_0>>p_{n}$ for even $n$.

However, this is not the case when $|r|<0.5$ where all the excited states seem to have a
contribution comparable to that of the ground state. This is clear in the insets of Fig.
\ref{fig:Fig2} where we plot $p_n/p_0$ vs $r$ for the corresponding values of drive parameters. 
The presence of a transition at $|r|=0.5$ is consistent with the phenomenon of PR in the
classical case i.e. Eq. \eqref{eq:eh}.
It is worth re-emphasizing here that in spite of this agreement with classical systems, the
phenomenon of PR in a quantum oscillator, especially the evolution of the
ground state, is starkly different due to the delocalized nature of the ground state 
wave function which sees changes in the potential even away from $x=0$.

\begin{figure}[H]
\centering
\ig[width=8.6cm]{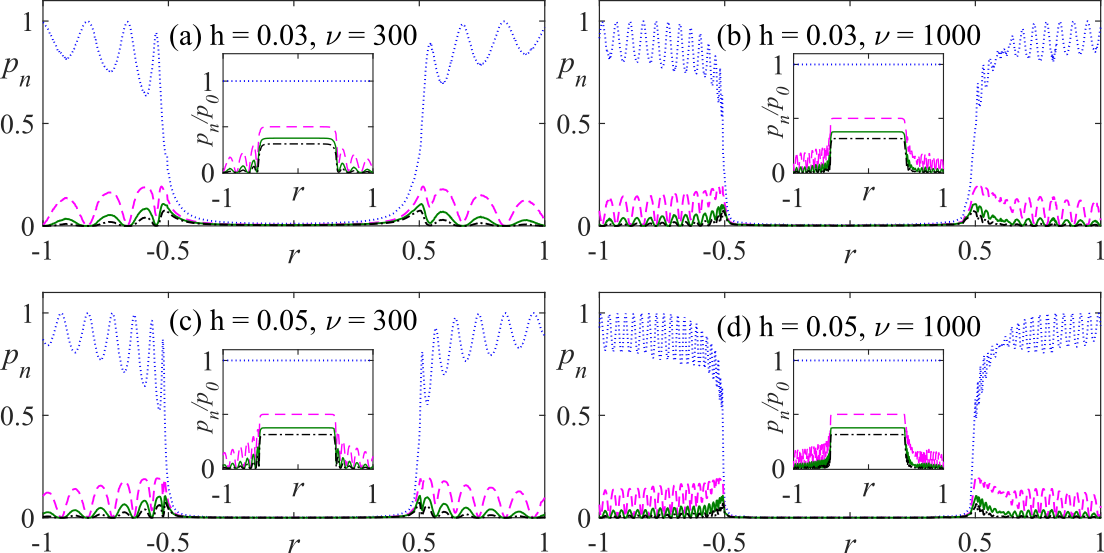}
\caption{The behavior of the probability $p_n = |\langle\Psi_n|\psi_0(\tau_f)\rangle|^2$ as a
function of $r = \bar\eps/h$ for different drive parameters. The blue, magenta, green and black
curves correspond to $n=0,2,4,6$ respectively. The transition close to $r = \pm 0.5$ becomes
sharper with increasing $h$ and $\nu$. The inset shows the ratio of $p_n/p_0$ for
$n=0,2,4,6$. In the region $|r|<0.5$ the quantum harmonic oscillator exhibits PR
since the evolved state has comparable weight in the ground state as well as in the higher
excited states. Outside this regime, the evolved state is primarily composed of the ground state
with the contribution from higher $n$ states being significantly lower.}
\label{fig:Fig2}
\end{figure}

{Fig.\ref{fig:Fig2} shows the probabilities for different energy eigenstates both 
outside the PR region ($|r|>0.5$) as well as inside it ($|r|<0.5$). For well
outside the PR region, we see that the probability of the excited states is much smaller than
that of the $n=0$ state. Inside the PR region the total probability is distributed
among several excited states, with $p_n$ being very small for every even $n$. This is seen from the
insets in Fig.\ref{fig:Fig2} where for each set of $(h,\nu)$ $p_n/p_0$ is plotted as a function of $r$
for $n=0,2,4,6$ as blue, magenta, green and black lines respectively.
The transition at $|r|=0.5$ is fairly sharp. 

A similar analysis is performed for the case when the initial state is the first excited state of the
unperturbed oscillator. In that case too, a sharp jump is seen in the probabilities. 
{{{\section{Excitation spectrum in the classical PR regime}}}}
In this regime, it is more useful to consider the ratio,
$p_n/p_0$. The reason for this is that in this regime, the total probability is distributed among so
many excited state that the individual contributions become very small. In the insets of Fig. \ref{fig:Fig2}
we notice that the probabilities of the excited states
are much closer to $p_0$ as compared to the region outside PR. The sharp nature of the transition at
$|r|=0.5$ also comes out very clearly in these plots. To highlight the fact that a significant
number of excited states are populated, although the probability $p_n$ decreases with increasing
$n$, we plot $ln(p_n/p_0)$ vs $ln(n)$ in Fig. \ref{fig:powerlaw}, for a fixed set of drive parameters.
Since this is a straight line, it is straightforward to conclude that in the PR region the probability
follows a power law behavior given by

\beq
p_n \propto n ^ \beta 
\eeq 
with $\beta$ being the slope of the straight line. We find that $\be<0$ in the resonance regime which
is consistent with the observation that the population of states decreases with increasing $n$.
{Further, the fact that $\beta > -1$ implies significant excitations to several higher
energy states.}} Moreover we find that for a fixed amplitude $h$ and drive cycles $\nu$, $\be$ is
found to be fairly insensitive to the value $r$ within the resonance regime.

If the power law dependence were exact, the probability conservation would be violated, due
to the non-converging nature of the series $\sum_{n=0}^\infty n^\beta$, since $\beta>-1$.
This implies that the assumption of a power law dependence is valid only up to a
finite value of $n$ and the behaviour deviates for larger $n$.

\begin{figure}[htb]
\centering
\ig[width=7.5cm]{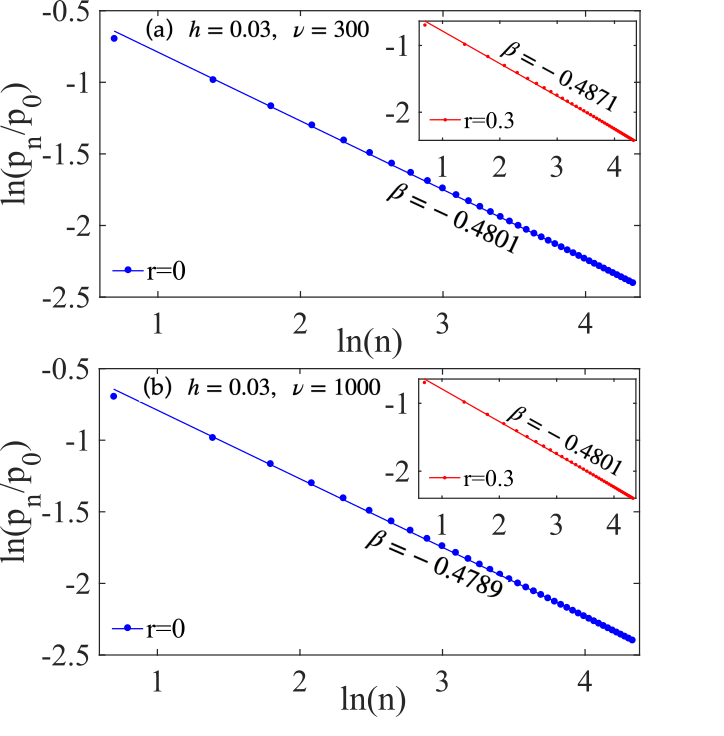}
\caption{Power law behavior in the PR regime ($|r|<0.5$) for different sets of drive parameters.
Since the log-log plot is linear, the population of states scales as a power law in the range of
$n$ shown here. Since a large number of states are populated, the value of $p_n$ for a fixed $n$
is very small in this regime, as can be seen from Fig. \ref{fig:Fig2}.}
\label{fig:powerlaw}
\end{figure}

{\section{Excitation spectrum outside the classical PR regime}}
{
{{The quantum effects outside the PR regime are shown in Fig. \ref{fig:Fig2}.
It is clear that in this region  ($|r|>0.5$)}}
the ground state significantly dominates over the higher energy states.
The transition probability is plotted as a semilog plot of $p_n$ {\it vs} $n$ in Fig. \ref{fig:exp}.
As is evident from the straight-line nature of the plot, the probability $p_n$ decays exponentially with
increasing (even) $n$ i.e.,
\beq
p_n \propto e^{\al n} \label{eq:pnexp}
\eeq
with $\al<0$ being the slope of the line in the semilog plot.
{This shows that the excitation to higher energy eigenstates decreases exponentially with $n$.
Therefore we conclude that, unlike the PR regime, here there is no significant excitation. This
fits into the classical result outside of the parametric regime where the amplitude does not
increase significantly.}
\begin{figure}[htb]
\centering
\ig[width=6cm]{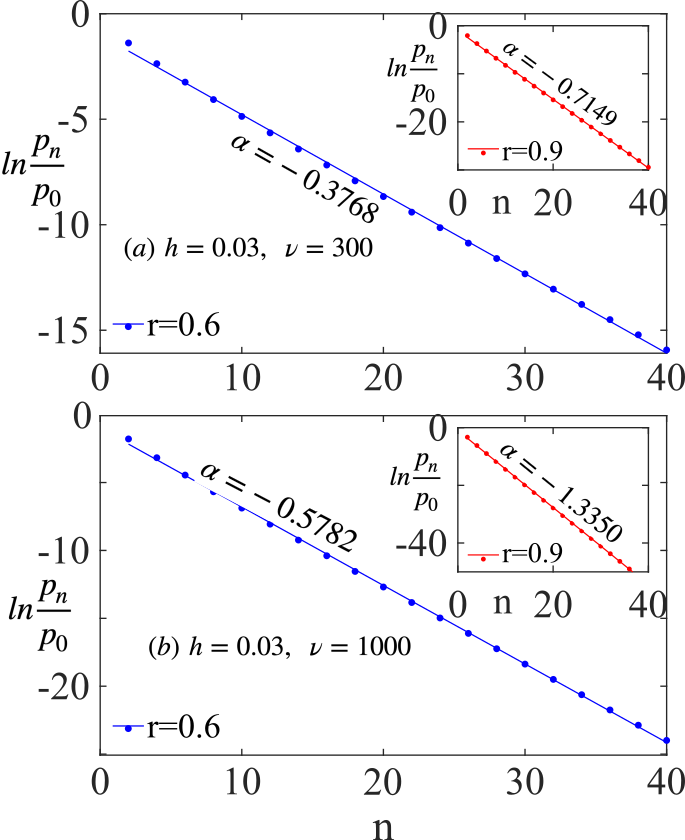}
\caption{Exponential decay of $p_n$ outside the PR regime $|r|>0.5$ for (a) $h=0.03, \nu = 300$
and (b) $h=0.03, \nu=1000$. The outsets correspond to $r=0.6$ while the inset has $r=0.9$ in each
of these subplots. The straight line trend of this semilog plot shows that the probability $p_n$
decays exponentially with increasing $n$ as in Eq. \eqref{eq:pnexp}. This means that outside the
parametric resonance regime, the ground state has the most significant contribution to the
time-evolved states and higher energy states have a much smaller role to play. This behaviour is seen 
for a wide range of drive parameters. However the exponent $\alpha$ now varies significantly as the
drive parameters are changed.}
\label{fig:exp}
\end{figure}
\begin{figure}[h!]
\ig[width=8cm]{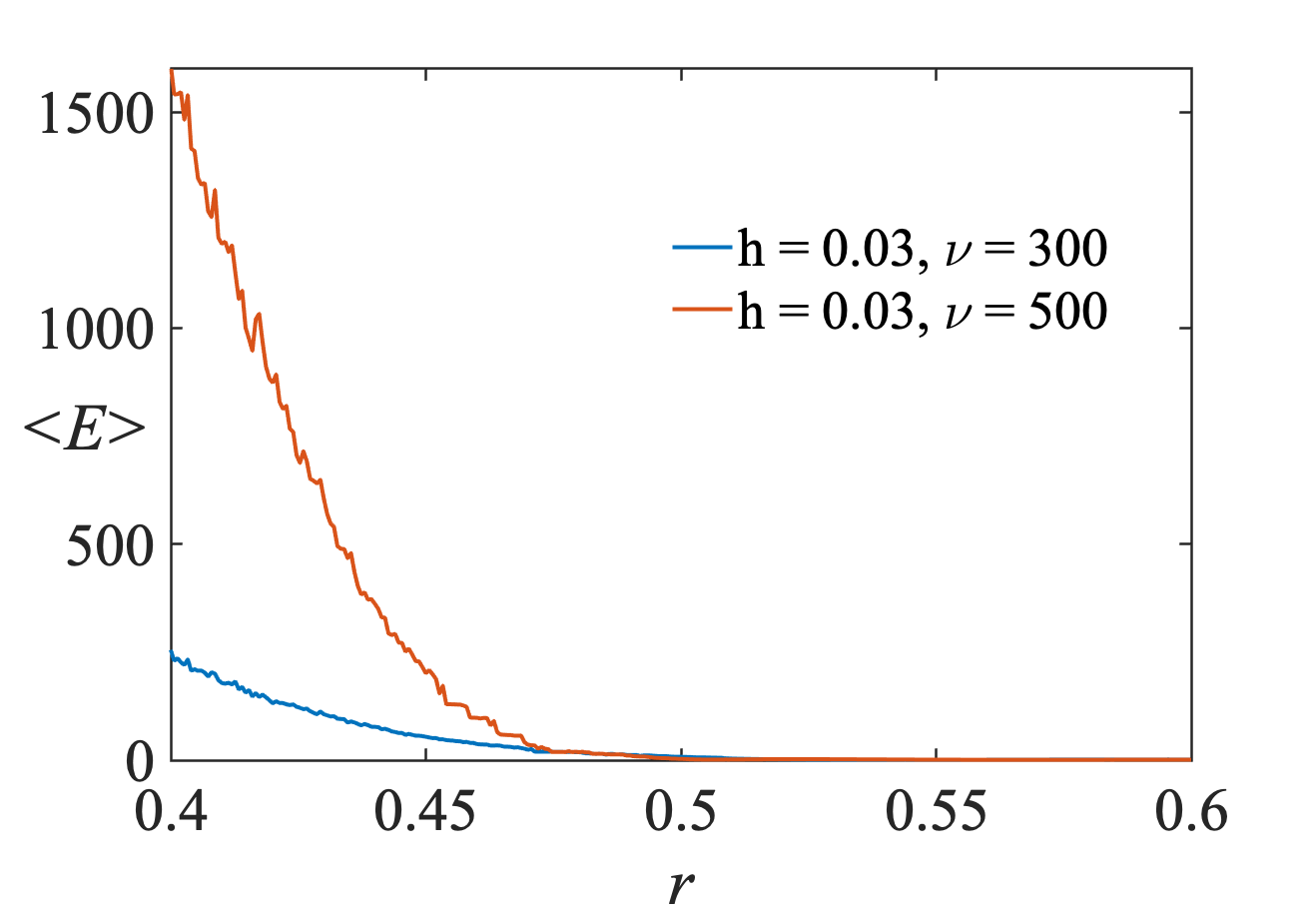}
\caption{The energy expectation value for a parametrically driven oscillator shows a 
sharp transition at $|r|=0.5$. In the PR regime, the energy absorbed by the oscillator is
very high. This is seen for two sets of drive parameters here.}
\label{fig:Hexp}
\end{figure}
\vspace{0.5cm}

\section{Energy absorption in the ground state}
An important aspect of resonance in any driven oscillator is]that when the resonance condition
is (almost) met the energy pumped into the oscillator is very high. To see this, we calculate
the expectation value of the Hamiltonian $\mathcal{H}_0$ at $\tau=\tau_f^+$, i.e.  $\la E\ra =
\langle\Psi(\xi,\tau_f)|\mathcal{H}_0|\Psi(\xi,\tau_f)\rangle$. Note that we use the unperturbed
Hamiltonian $\mathcal{H}_0$ here, because the drive is switched off at $\tau=\tau_f$. 
Figure \ref{fig:Hexp} for {(a) $h=0.03, \nu=300$ and (b) $h=0.03, \nu=500$} shows that the energy
absorption is very high well within in PR regime and there is a rapid drop to negligible value of
$\la E \ra$ as we move outside this regime. Additionally, this transition becomes
more pronounced with higher number of drive cycles $\nu$.

This observation can be explained as follows. The energy expectation value $\la E\ra$ can be 
expressed in terms of the
energy eigenvalues $E_n$ of the unperturbed oscillator as
\beq \la E \ra = \sum_{n} p_n E_n\eeq
where $p_n$s are the probabilities of Eq. \eqref{eq:pndef}.
Since we have shown in Figs. \ref{fig:powerlaw} and \ref{fig:exp} that there is a transition
in the nature of $p_n$ vs $n$ in and outside of the PR regime, it immediately follows that
$\la E\ra$ is expected to behave differently in the two regions.
The most important point to note here is that this energy pumping happens even in the ground
state of the oscillator which is in complete contrast with the classical parametric resonance,
where the minimum energy configuration does not absorb energy via this drive.

\section{Summary}
In this work, we have investigated the evolution of the ground state of a harmonic oscillator when
subjected to a perturbation that modulates the natural frequency of the oscillator. In particular, we
study the special case when the classical parametric resonance conditions are met. Since the classical
``ground state" (minimum energy configuration) does not evolve under these conditions, the effects
discussed in this paper are those which arise purely due to the quantum nature of the system.
It is shown that given the time-dependence of the form Eq. \eqref{eq:ft},  there is sharp transition
in behavior at $r=|0.5|$, which is seen in both the probability $p_n$ and well as the energy
expectation value $\la E\ra$. The transition probability $p_n$ of the time-evolved ground state always decreases
monotonically with even $n$. While within the PR regime this
follows a weak power law behavior, outside this regime, an exponentially decaying trend is seen. The
weak power-law nature ensures significant transition probability into higher excited states. The fact that
this is seen for a quantum ground state (in contrast to the minimum energy configuration of classical case)
is due to the delocalized nature  of the  ground state of the quantum oscillator. 

\section {Acknowledgements}
The author thanks Diptiman Sen for useful insights and comments on the manuscript.

\bibliography{refs}
\end{document}